\documentclass{ws-rv9x6-mod}
\usepackage{ws-rv-van}% default - numbered citation/references
\usepackage[normalem]{ulem}
\usepackage[utf8]{inputenc} 

\usepackage{hyperref}

\newcommand{\tvec}[1]{\boldsymbol{#1}}
\renewcommand{\vec}[1]{\boldsymbol{#1}}
\newcommand{\pr}[2]{{}^{#1}\! #2}
\newcommand{\jpsi}{J\mskip -2mu/\mskip -0.5mu\Psi}

\newcommand{\gev}{\,\operatorname{GeV}}

\newcommand{\fm}{\operatorname{fm}}

\begin{document}
\chapter[DPS correlations]{Parton correlations in double parton scattering}

\author[T. Kasemets]{T. Kasemets}
\address{PRISMA Cluster of Excellence \& Mainz Institute for Theoretical Physics Johannes Gutenberg University, 55099 Mainz, Germany}
\author[T. Kasemets and S. Scopetta]{S. Scopetta}
\address{
Department of Physics and Geology, University of Perugia
and INFN, Sezione di Perugia, Via A. Pascoli, I-06123, Perugia, Italy}
\begin{abstract}
Double parton scattering events are directly sensitive 
to the correlations between
two partons inside a proton
and can answer fundamental questions on the connections
between the proton constituents. 
In this chapter, the different types
of possible correlations,
our present knowledge of them, and the
processes where they are likely to be important,
are introduced and explained. 
The increasing integrated
luminosity at the LHC
and the refinements of the theory of 
double parton scattering, 
lead to interesting prospects for measuring, 
or severely constraining, two-parton correlations
in the near future.
\end{abstract}
\body
%%%%%%%%%%%%%%%%%%%%%%%%%%%%%%%%%%%
\section{Introduction}
%%%%%%%%%%%%%%%%%%%%%%%%%%%%%%%%%%%
The study
of double parton scattering (DPS) events can open up a window to see, 
for the first time, 
how the constituents of the proton are connected to each other. 
The correlations between the properties of 
two partons in one proton can be directly probed,
measuring how two partons inside the proton affect 
one another. So far, only indirect
tests of these correlations have been possible,
studying for example, by means of electromagnetic interactions,
how the collective 
behavior of the constituents sums up
to give the proton spin.

This allows us to answer questions such as: 
How does the probability to find one parton in a certain spin state 
affect the probability to find the second parton in the same spin state? 
In this chapter we 
will look at two quarks or gluons inside the proton, 
explain the different ways they can be connected to one another, 
describe the state-of-the-art of the field
as well as the perspectives for future studies of these correlations.
This will be possible in
processes where DPS forms a major contribution, such as 
same-sign double-$W$ production.

Assuming factorization (see Ref. \refcite{Diehl:2017wew}),
the DPS cross section for the production of 
final states $A$ and $B$ {takes the form
\cite{Diehl:2017wew,Paver:1982yp} 
\begin{equation}
d \sigma^{AB}_{DPS} = {m \over 2} 
\sum_{abcd,R} \int d^2 \tvec{y} \,
\pr{R}{F}_{ac}(x_1,x_2,\tvec{y})
\pr{R}{F}_{bd}(x_3,x_4,\tvec{y}) \, 
d \pr{R}{\hat \sigma}_{ab}^A \, d \pr{R}{\hat 
\sigma}_{cd}^B \,,
\label{uno}
\end{equation}
where $m=1$
if $A$ = $B$, $m=2$ otherwise, $R$ denotes the different possible 
color representations and 
$a,b,c,d$ label simultaneously the species (parton-type and flavor) 
and polarization of the partons contributing
to the production of the final states. In Eq. \eqref{uno}, 
$d \hat{\sigma}$ represents the differential partonic cross section 
(for example, differential in the rapidities of the 
produced particles).  
The functions $F$
are the double parton distributions (dPDFs), 
encoding the probability to find the two interacting 
partons, with longitudinal 
fractional momenta $x_1,x_2$ at a relative transverse distance
${\tvec{y}}$ inside the proton. 
They depend additionally
on factorization scales $\mu_{A(B)}$, and for $R \neq 1$, on
a rapidity scale \cite{Diehl:2017wew}. If extracted from data, 
as noticed a long time ago \cite{Calucci:1999yz},
dPDFs would offer for the first time 
the opportunity to investigate two-parton correlations.
This would be a two-body property,
carrying information
which is
different and
complementary to that encoded in 
one-body distributions, such as generalized parton distributions 
(GPDs) \cite{Diehl:2015uka}. 
This is illustrated in figure~\ref{fig:two_body}.

For cross sections differential also in the net 
transverse momenta of each of the
two hard interactions, the dPDFs are replaced in the factorization theorem 
by the 
double transverse momentum dependent parton distributions (dTMDs).
{These distributions 
depend on two additional transverse vectors 
and allow for
a number of further correlations, for example between the spin and transverse
momenta of the partons. They are interesting also from a
more theoretical point of view, with the rich color structure in combination with the non-trivial dependence
on the soft gluon exchanges \cite{Buffing:2017mqm}.}
In the region where the two net transverse momenta are small, 
DPS and single parton scattering (SPS)
 both contribute to the cross section at the same power, which makes 
it promising for DPS extractions.
However, for simplicity 
we will focus on the dPDFs during the rest of this chapter.

In the following, we will have a closer look at what is currently known about the
different correlations, describe the effects 
expected in cross sections and the prospects 
for their measurement. The chapter is structured as follows: 
In the next section, we will look at the correlations 
between the kinematic variables $x_i$ and $\tvec{y}$ of the dPDFs. 
In section 3, we will focus instead on the correlations 
between color, spin, flavor and fermion number of 
the two partons. 
In section 4 we will summarize and give an outlook to what we 
consider are the most promising future directions.

\section{Kinematic correlations}

{As stressed in the introduction, the two-body information
encoded in dPDFs is different and complementary to that
described by one-body parton distributions. 
Nevertheless, a connection between dPDFs, 
presently largely unknown, and
one-body quantities can be obtained by making
a number of assumptions on the dPDFs.
First, all color representations
different from the color singlet are neglected 
(i.e., only $R=1$ is considered),
together with all possible
correlations between spins, flavors and fermion-numbers.
Thereafter, correlations between $x_1$ and $x_2$
are neglected. The dPDFs then take the form
\begin{eqnarray}
\label{fact}
F_{{jk}}(x_1,x_2,\vec{y})  = 
\int d^2 \vec{b} F_{j} (x_{1}, \vec{b} +
 \vec{y} )
F_k(x_{2},\vec{b} )~,
\end{eqnarray}
where $F_{i}(x, \vec{b})$ is a parton
distribution dependent on the impact parameter 
$\vec{b}$, the transverse distance of
the parton from the transverse center of mass of the hadron
\cite{Diehl:2015uka}.
This function is the Fourier transform
of a GPD in a process where
the momentum transfer is transverse.
Neglecting moreover correlations between $x_1$, $x_2$ and 
$\vec{b}$, one can write
\begin{eqnarray}
\label{factb}
F_{i}(x, \vec{b} )=f_{i}(x)G(\vec{b})~,
\end{eqnarray}
where $f_{i}{(x)}$ is a parton distribution 
function (PDF) and the transverse profile
$G(\vec{b})$ has been assumed to be equal 
for all parton species. 
One should notice that 
Eq. (\ref{factb}) has been
found to fail in all model calculations of GPDs (see, e.g.,
Refs.
\cite{Scopetta:2002xq,Boffi:2002yy}),
as well as in the first analyses of  
data from deeply virtual Compton scattering \cite{Dupre:2016mai}.
The assumptions 
described above are often used to infer
properties of dPDFs from those of single particle distributions.
The relations Eqs. (\ref{fact}) and (\ref{factb}) have been
introduced and critically discussed, in a mean field approach, 
in Refs. \cite{Blok:2010ge,Blok:2011bu}.
%\cite{Blok:2017alw}

\begin{figure}
\includegraphics[width=10.cm]{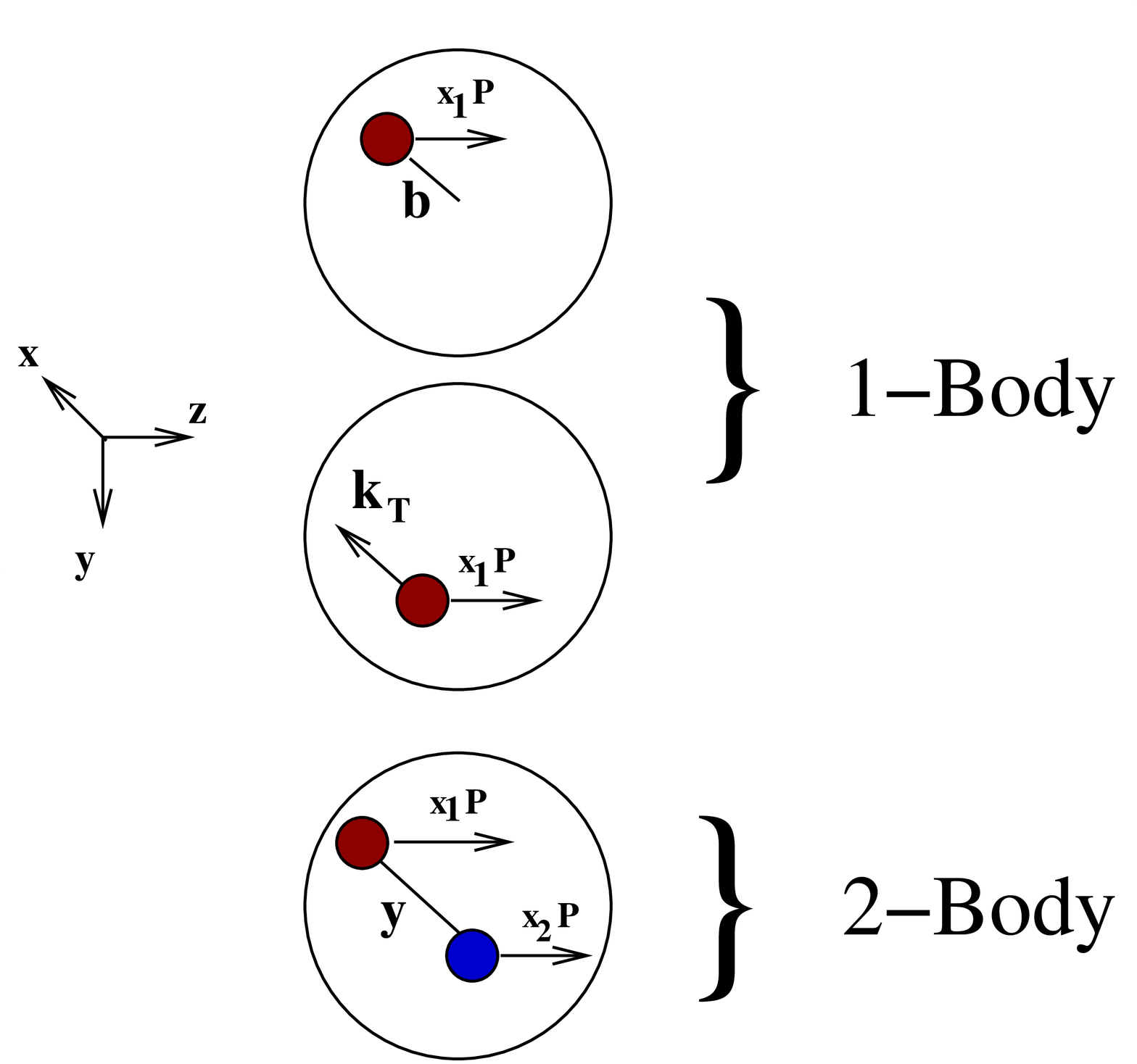}
\caption{From top to bottom: a pictorial representation of an impact
parameter dependent parton distribution, i.e. the Fourier transform
of a GPD when the momentum transfer is purely transverse;
a transverse momentum dependent parton distribution (TMD); 
a dPDF, which, at variance with the two previous cases, is a 
two-body distribution.}
\label{fig:two_body}
\end{figure}
Since dPDFs are largely unknown, and only
sum rules relating them to PDFs are available
\cite{Gaunt:2009re,Gaunt:thesis,Blok:2013bpa,
Ceccopieri:2014ufa,Plossl:2017wjw},
model calculations can be very useful and have been performed. 
Models 
are usually developed at low energy, but are able 
to reproduce some relevant features of nucleon parton 
structure.
Since in models the number of degrees of freedom
is fixed,
they can be predictive in particular in the valence region,
at $x$ larger than, say, 0.1.
In such model calculations, the factorized structures in 
Eqs. (\ref{fact}) and (\ref{factb}) do not arise.
Relevant correlations between $x_1$ and $x_2$,
violating Eq. (\ref{fact}),
and between $x_1,x_2$ and $\vec b$, violating Eq. (\ref{factb}),
have been found in the valence region in a variety
of approaches. This result was obtained, for example, in
a modified version of the simplest bag model
\cite{Chang:2012nw}, in constituent
quark models \cite{Rinaldi:2013vpa,
Rinaldi:2014ddl} in a valon model with QCD evolution
\cite{Broniowski:2013xba,
Broniowski:2016trx}
and in dressed
quark models \cite{Kasemets:2016nio}.
\begin{figure}
\includegraphics[width=1.2\textwidth]{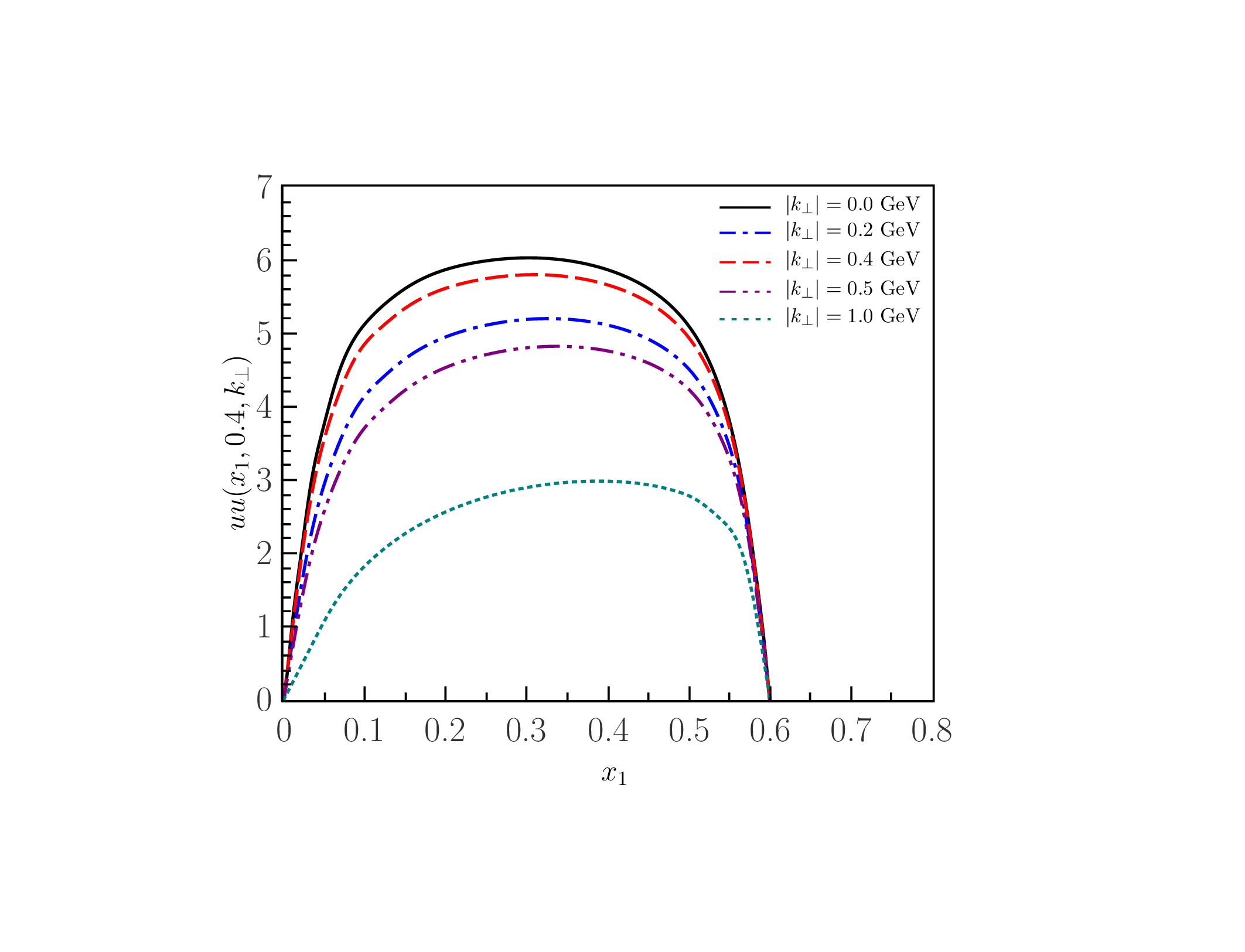}
\vskip -1.8cm
\caption{The distribution $uu(x_1,x_2,\vec{k}_\perp)$, Fourier
transform of the dPDF $F_{uu}(x_1,x_2,\vec{b})$, for the proton,
for $x_2$=0.4, according to the LF model calculation
of Ref. \refcite{Rinaldi:2014ddl}, at the low
momentum scale of the model.
If it were possible to factorize the dependence on the longitudinal
momenta $x_1,x_2$ and that on the transverse variable $\vec{k}_\perp$,
the distributions would have the same symmetric 
shape for the different values of 
$\vec{k}_\perp$.}
\label{fig:kperp}
\end{figure}

In particular, in Ref. \refcite{Rinaldi:2014ddl}
a light-front (LF) 
Poincar\'e covariant approach,
reproducing the essential sum rules of dPDFs
without ad hoc assumptions and containing
natural two-parton correlations, has been described.
An example of the information that model calculations
can provide is shown in Fig. \ref{fig:kperp}, where
the effect of the breaking of the factorization between
longitudinal and transverse variables is emphasized.

It is crucial to explore 
if the breaking of the properties,
Eqs. (\ref{fact}) and (\ref{factb}), found
in the valence region, survive at LHC kinematics, 
dominated by low-$x$ and high energy scales.
As a matter of fact, model estimates are valid in general
at a low scale
$Q_0$, the so-called hadronic scale.
The results of the calculations
should therefore be evolved using perturbative QCD (pQCD)
in order to compare them with data taken at a momentum scale
$Q>Q_0$, according to a well established procedure, proposed already
in Refs. \refcite{Parisi:1976fz,Jaffe:1980ti}.
The evolution of dPDFs has been 
studied for a long time. 
The first studies were performed in the late '70s/early '80s
\cite{Kirschner:1979im,Shelest:1982dg} with much theoretical 
progress being made in recent years -- for a detailed discussion 
on this topic we invite the reader to look at Ref. \refcite{Diehl:2017wew},
{\refcite{Blok:2017alw} and references there in}.
One should notice that, even if a factorized structure of the dPDF
were valid at a given scale, the different evolution properties
of dPDFs and PDFs would break it at a different scale,
generating perturbative correlations. 
These correlations
have been discussed in a largely model-independent way in Ref. 
\refcite{Diehl:2014vaa}{, incorporating the homogeneous evolution equations}. 
The evolution tends to pull the average transverse separation in quark 
and gluon distributions towards a common value, 
but this is a relatively slow process and differences can remain up 
to high scales. Similarly, correlations between the momentum fractions 
and the transverse separation present at a low scale can remain in 
large scale processes, as described here below. 

The interplay of perturbative and non-perturbative
correlations between
different kind of partons has been described also using 
homogeneous QCD evolution
applied to the results of the correlated LF model \cite{Rinaldi:2016jvu}.
It was found that their effect tends
to be washed out at low- $x$ for the valence,
flavor non-singlet distributions, while they can affect
singlet distributions in a sizable way.
This different behavior can be understood
in terms of a delicate interference of non-perturbative correlations, 
generated by the dynamics of the model, and perturbative ones,
generated by the model independent evolution procedure.

Concerning the correlation between the $\vec{y}$ 
and $x_1, x_2$ dependences in dPDFs, 
some qualitative understanding can be inferred 
from studies of hard exclusive processes, 
involving 
$f_i(x,\vec{b})$ of a single parton inside the proton.  In particular,
measurements of $\gamma p\to \jpsi p$ at HERA
\cite{Chekanov:2002xi,Aktas:2005xu} indicate a logarithmic dependence
$\langle \vec{b}^2 \rangle = \text{const} + 4 \alpha' \log(1/x)$ with
$\alpha' \approx 0.15 \gev^{-2} = (0.08 \fm)^2$ for gluons with $x \simeq
10^{-3}$.  Studies of nucleon form factors 
\cite{
Diehl:2013sca
} and
calculations of Mellin moments $\int dx\; x^n f_i(x,\vec{b})$ with
$n=0,1,2$ in lattice QCD \cite{Hagler:2009ni} indicate that for $x$ above
$0.1$ the decrease of $\langle \vec{b}^2 \rangle$ with $x$ is even
stronger.  Although this is one-body
information,
one could wonder whether the
correlations between the $\tvec{y}$ 
dependence and $x_1, x_2$ in double
parton distributions could follow the behavior
of the one-body quantity,
with the $\vec{b}$ distribution becoming more
narrow with increasing $x$.
If this is the case, 
important consequences could be expected
for multiparton interactions
\cite{Frankfurt:2003td}.
The production
of hard final states requires relatively
large momentum fractions of the partons entering the corresponding hard
interaction. This would favor
small values of $\vec{b}$, which is the
transverse distance of the parton from the transverse
center of the proton. 
The collision would therefore be rather central
and thus the 
transverse interaction area for the colliding protons
would be rather large,
a fact which in turn favors
additional interactions.

Such
correlations may have a sizable impact, e.g., on the underlying
event activity in $Z$ production,
as shown in a study with Pythia 8
\cite{Corke:2011yy}.

%%%%%%%%%%%%%%%%%%%%%%%%%%%%%%%%%%%%%%%%%%%%%%%%%
\section{Quantum-number correlations}

Two partons inside a single proton can have their quantum numbers 
correlated. Perhaps the most straightforward example comes from the 
valence sector of the proton. If we, for one interaction, 
extract one valence up quark from the proton, it is natural to expect 
that the chance to find another {valence up} quark in the proton is reduced. 
It seems reasonable to expect such effects to be sizable at relatively large 
momentum fractions and to reduce as the density of partons increases 
towards small momentum fractions. {This phenomenon} naturally fits into the 
dPDFs, $F_{ab}$, of two partons $a$ and $b$ 
inside a proton. 

We will focus here on another type of correlation and interference which 
occur{s} at the quantum level, and for which we reserve the label 
{\it quantum-number correlations}. This includes correlations and 
interferences in color, spin, flavor and fermion number
\cite{Manohar:2012jr, Diehl:2011tt, Mekhfi:1985dv}. 
Understanding how this occurs in double parton scattering, 
but not in single parton scattering, is not complicated. 
From a diagram {such as the one} in figure~\ref{fig:DVBprod}, we can see that 
two quarks {\it leave} the right{-}moving proton 
(represented by the lower green ellipse) on the left side of the final-state 
cut  and two quarks {\it return} to the proton on the right side of the cut. 
The quantum numbers of the two quarks in the amplitude have to sum up to 
the quantum numbers in the conjugate amplitude, which still leaves 
room for the two quarks in the amplitude to individually have different 
quantum numbers {from} their partner{s} in the conjugate {amplitude}.
\begin{figure}
\centerline{\includegraphics[width=0.5\textwidth]{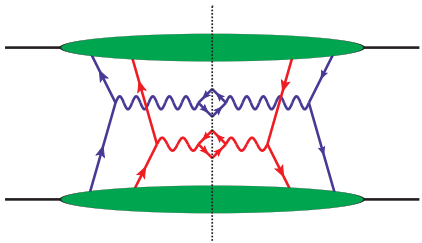}}
\caption{Double vector boson production. In contrast to single parton 
scattering, only the {\it sum} of the quantum numbers of the partons 
leaving the protons on the left and {returning on the} right hand side of the final-state 
cut have to match.}
\label{fig:DVBprod}
\end{figure}

In particular, this allows for quantum number interferences, which is another 
way of viewing the correlations. If we take color {as example} 
(even though, as we will see, it might not have the largest impact), 
and couple each parton in the amplitude with its partner in the conjugate 
amplitude (i.e. {parton} with the same longitudinal momentum fraction $x_i$) 
we have two possible combinations{:} 
$3 \otimes \bar{3} = 1 \oplus 8${.} 
{R}epeating this with the other pair we obtain
\begin{align}
(3 \otimes \bar{3})\otimes (3 \otimes \bar{3}) = (1\otimes 1) \oplus (1 \otimes 8) \oplus (8 \otimes 1) \oplus (8 \otimes 8) = 1 \oplus 1 \oplus ...
\end{align}
where the "..." {refer to} combinations {that do} not produce a total color 
singlet. The requirement that the sum of the quantum numbers on the left 
and right side of the final-state cut have to be equal amounts to the 
requirement that when coupling all {four} partons, we need to obtain a color 
singlet. We therefore see that for the quark case we can obtain the 
singlet in two ways{: either} by coupling two individual color singlet 
pairs or by 
coupling two color octet pairs. This results in two independent 
double quark distributions for the two color states in Eq.~\ref{uno}, labeled as $\pr{R}{F}$ 
with $R=1,8$. In the cross section, both distributions contribute and 
color{-}singlet production is proportional to 
$\pr{1}{F}\pr{1}{F}+\pr{8}{F}\pr{8}{F}$ (with the normalization of the 
distributions as in \cite{Diehl:2011yj}). The color{-}octet {term}
has hard interactions with color interference{s} between the amplitude and 
conjugate, i.e. {it} is a genuine quantum effect which can never 
appear in a single hard scattering. Under the 
assumption of zero correlations between the two hard interactions, no such 
interference could take place and the octet distributions would vanish. 

Similar to the color, also the spin of the two partons can be correlated and 
give rise to a large number of different {polarized} dPDFs. 
There can be interferences in flavor, for example between up and down quarks 
in double-$W$ boson production. This type of interference is illustrated by the
diagrams in figure~\ref{fig:interf-W}. 
\begin{figure}
\centerline{\includegraphics[width=0.6\textwidth]{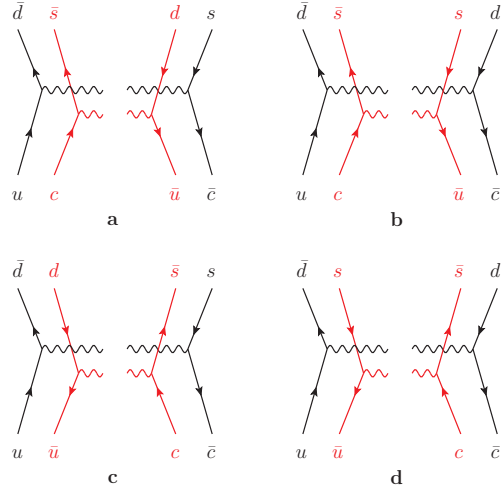}}
\caption{Flavor interference in double $W$ production. Two possible processes are shown for $W^+W^+$ production in (a,b), and for $W^+W^-$ production in (c,d). Figure from Ref. \refcite{Kasemets:2012pr}. $q$ and $\bar{q}$ labels partons corresponding to a quark field or a conjugate quark field in the relevant dPDF. Graphs (b) and (d) have flavor interference only for the proton at the bottom, while graphs (a) and (c) come with flavor interference distributions for both protons.}
\label{fig:interf-W}
\end{figure}
Furthermore, there can be interference in fermion number between quarks, 
antiquarks and gluons as examplified in figure~\ref{fig:num-int}.
\begin{figure}
\centerline{\includegraphics[width=0.35\textwidth]{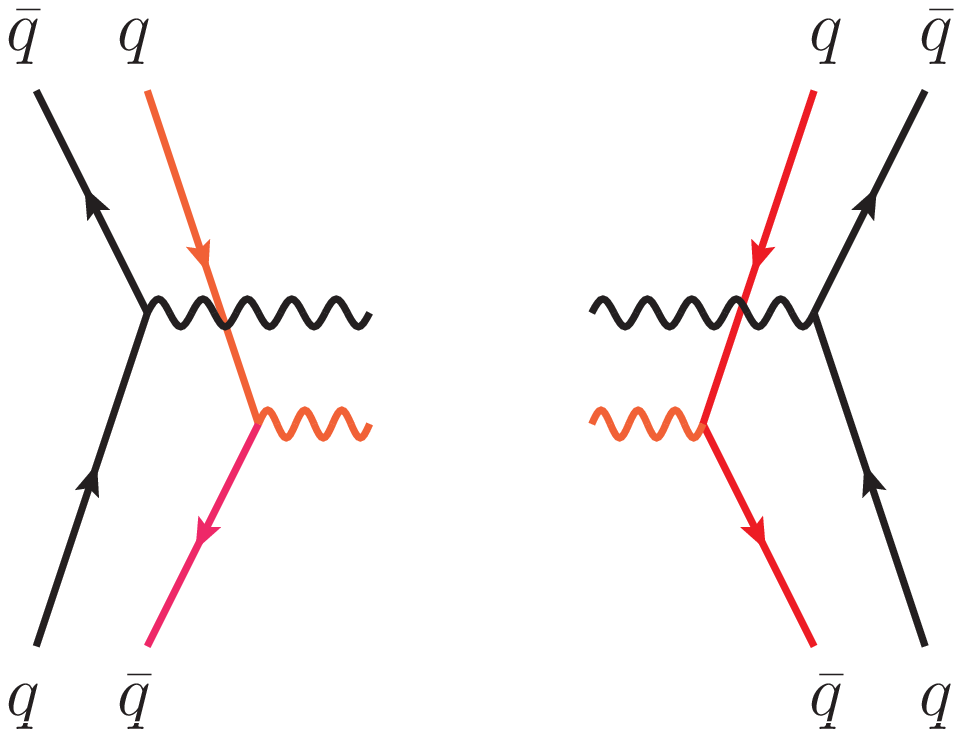} \;
\includegraphics[width=0.35\textwidth]{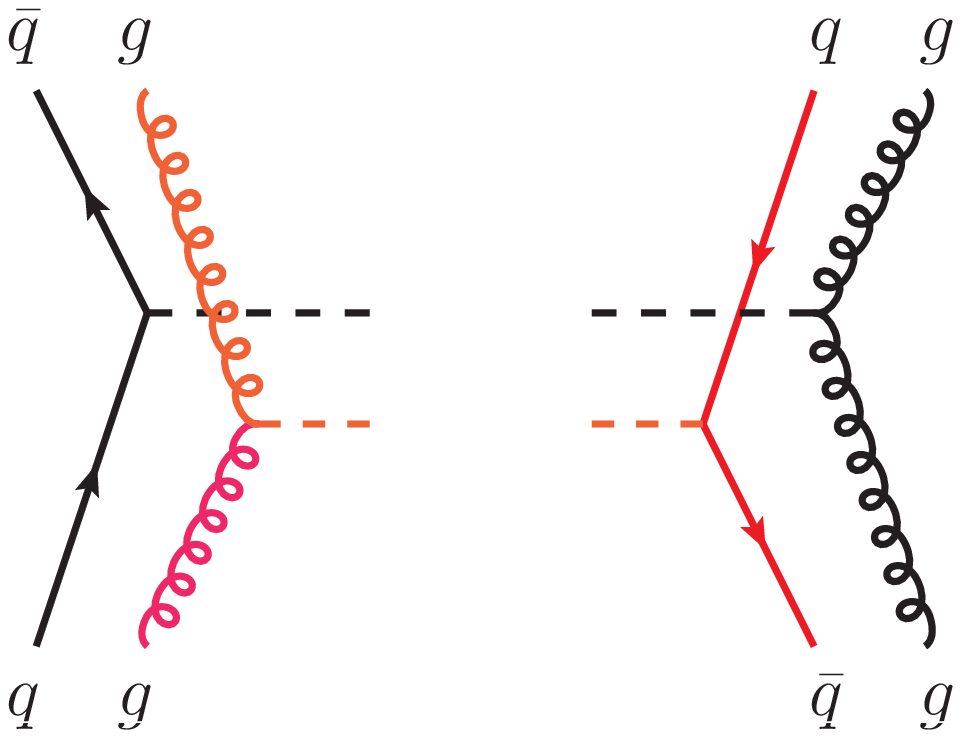}}
\caption{Fermion number interference examples for double Drell-Yan (left) and double Higgs production (right). $q$ and $\bar{q}$ labels partons corresponding to a quark field or a conjugate quark field in the relevant dPDF and $g$ labels a gluon field.}
\label{fig:num-int}
\end{figure}
It is interesting to note, that 
spin correlations lead{ing} to distributions 
of transversely polarized quarks and linearly polarized gluons
have a 
rather unique signature. They induce a dependence on the azimuthal angle 
(for example between the $Z$-boson decay planes) and lead to azimuthal {spin} 
asymmetries in unpolarized proton scattering \cite{Kasemets:2012pr}. {It is important to realize that experimental extractions of DPS signals are based on Monte Carlo generators which assume a flat azimuthal distributions, which might no longer be true in the presence of correlations. }

{The result of all the correlations is} a flora of independent double parton distributions, of which we have little knowledge and {no} experimental extractions. One might question what predictive power we have, and can hope to obtain. The answer to this question leads us into a discussion of what we know about the different correlation effects, when they are likely to play an important role and when we believe they can be safely neglected. The information available to this end come{s} from two main categories of studies. The first studies the distributions in different types of hadron models, or deriv{es} theoretical bounds, and attempts to quantify the size of the correlation{s}. The second examines how the perturbatively calculable evolution of these distributions influence{s} their shapes and sizes. 

\subsection{Models and bounds}
There are a couple of different hadron model calculations which consider 
different quantum-number correlations. Focus has been on the polarization of the partons{ (apart from the kinematic correlations already discussed) 
and large correlations have been observed}.
For quark and antiquark distributions, large spin correlations {were} found in the MIT bag model \cite{Chang:2012nw} and in light-front constituent quark models \cite{Rinaldi:2014ddl}. The domain of validity of these models is principally the region of large momentum fractions, and thus they serve best as initial conditions to the double DGLAP evolution equations. This was done in \cite{Rinaldi:2014ddl} with the observation that the spin correlations are sizable even after evolution. Within a dressed-quark model of the mixed quark-gluon distributions, the spin 
correlations
{were} observed to be large for certain polarization types, such as {two} longitudinally polarized {partons} and {the combination of} a transversely polarized quark and an unpolarized gluon \cite{Kasemets:2016nio}. In addition to model calculation{s}, {theoretical} upper bounds on the correlations, including spin, flavor, fermion number and color, have been derived from the probability interpretation (or positivity) of {dPDFs} \cite{Kasemets:2014yna, Diehl:2013mla}.

\subsection{Evolution}
The {d}PDFs evolve according to a double ladder version of the DGLAP evolution equations, i.e. a double DGLAP evolution \cite{Diehl:2017wew}. Cross talk between the ladders {is} suppressed by the large distance $\tvec{y}$ separating the two partons, which is typically of the size of the proton. The evolution starts at a scale of the order of 
$1/|\tvec{y}|$ and evolves up to the scale of the respective hard interaction\cite{Diehl:2017kgu}. This evolution generically leads to a reduction of the correlation{s} between the two partons and decreases the importance of the two interference/correlation {d}PDFs. However, the rate at which this occurs varies significantly for the different types of correlations and the momentum fractions of the partons.

If we allow for a slight oversimplification{, the current} state of
knowledge can be summarized in a short paragraph: The color correlations are 
{Sudakov} suppressed and expected to be small in large{-}scale processes\cite{Manohar:2012jr,Mekhfi:1988kj,Diehl:2017wew}. This can be understood from the fact that those correlations require color information {to} travel {over the} large distance $\tvec{y}$ inside the proton. Therefore, for processes above $Q_i^2 \sim 100~\gev^2$ they are expected to play a minor role. 

{Gluon polarizations at low momentum fractions (where DPS is most relevant) are also quite rapidly suppressed through the evolution. This suppression can be understood from the gluon splitting kernels: The unpolarized gluon splitting kernel at small $x$ goes as $1/x$, leading to the large increase of the gluon density for small momentum fractions (as is well known from single parton distributions). The polarized splitting kernels on the other hand go as $x^0$ for longitudinal polarization and $x$ for linearly polarized gluons. The quark polarizations on the other hand can remain sizable up to high scales\cite{Diehl:2014vaa}. Figure~\ref{fig:pol-evo} shows two examples of the suppression for the most suppressed gluon polarization and the least suppressed quark polarization, starting with maximal polarization (i.e. polarized equal to unpolarized) at the input scale of 1~GeV.} Fermion-number interference is expected to be small at large scales, since the interference distributions do not mix with the gluon {distributions} (which drives {the} evolution at small to moderate $x$) under leading order evolution and always involve color interference. Flavor interference on the other hand is still relatively unexplored, but also does not mix with the gluon distributions. 

\begin{figure}
\centerline{
\includegraphics[width=0.48\textwidth]{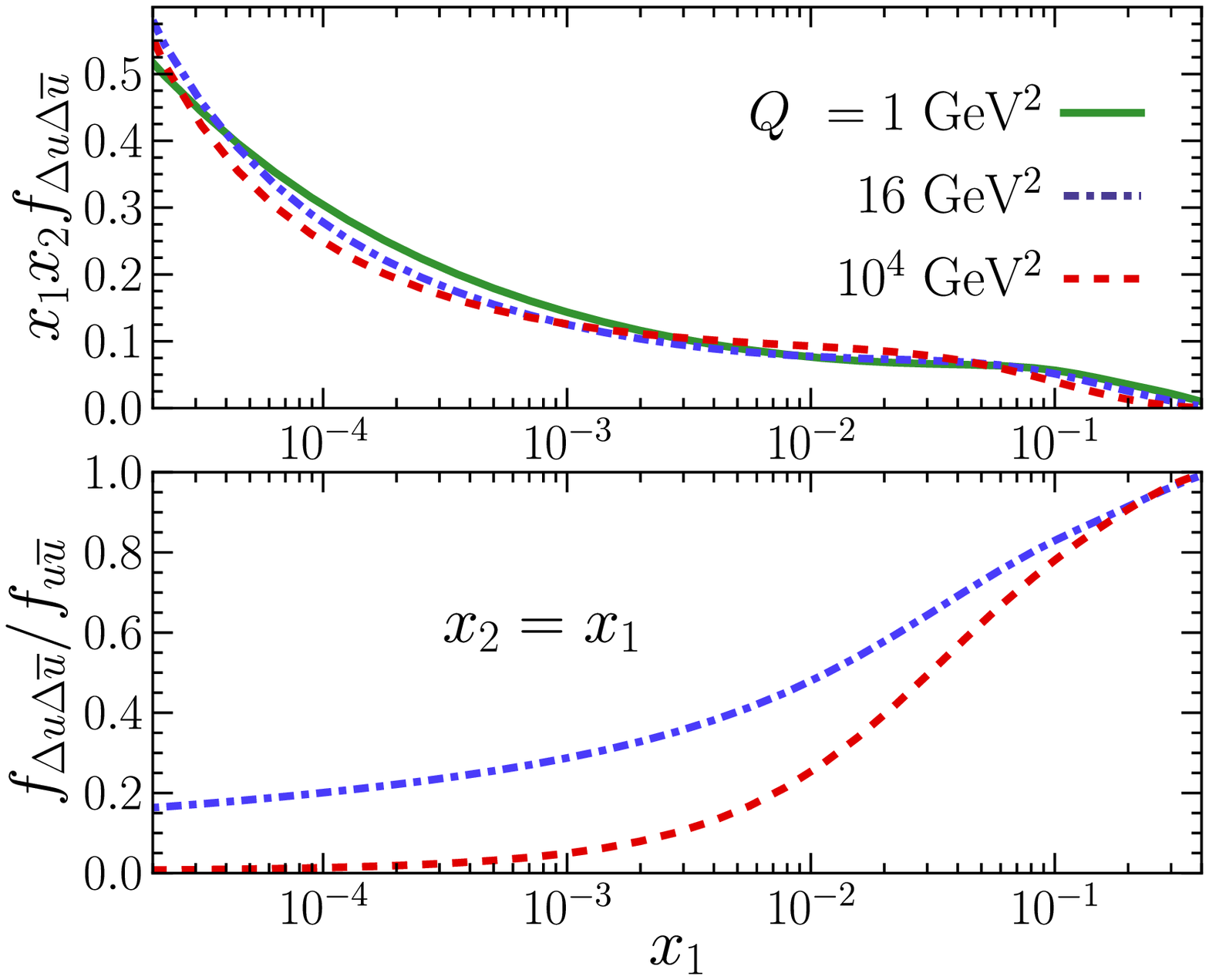} \;
\includegraphics[width=0.48\textwidth]{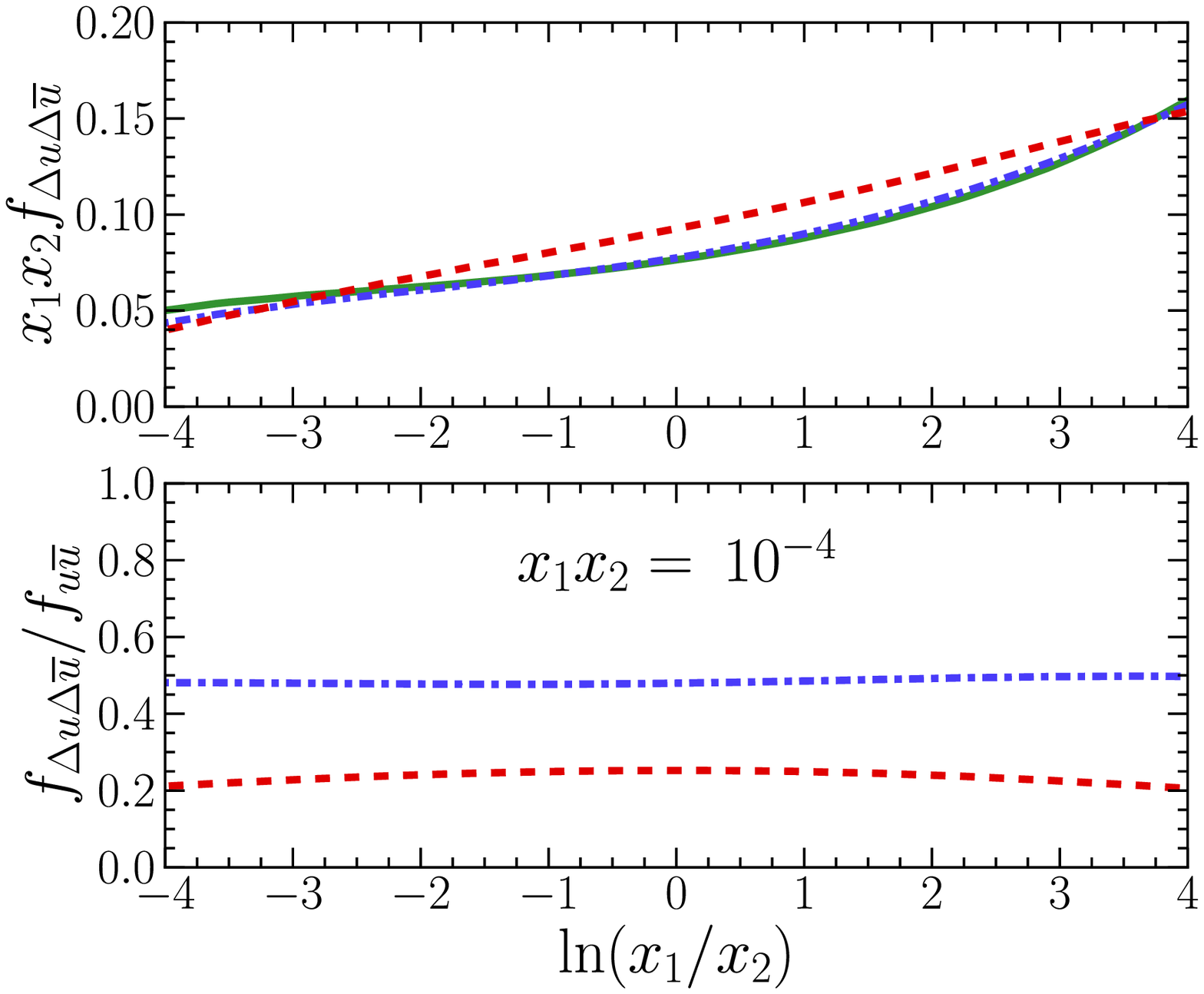}}
\centerline{\includegraphics[width=0.496\textwidth]{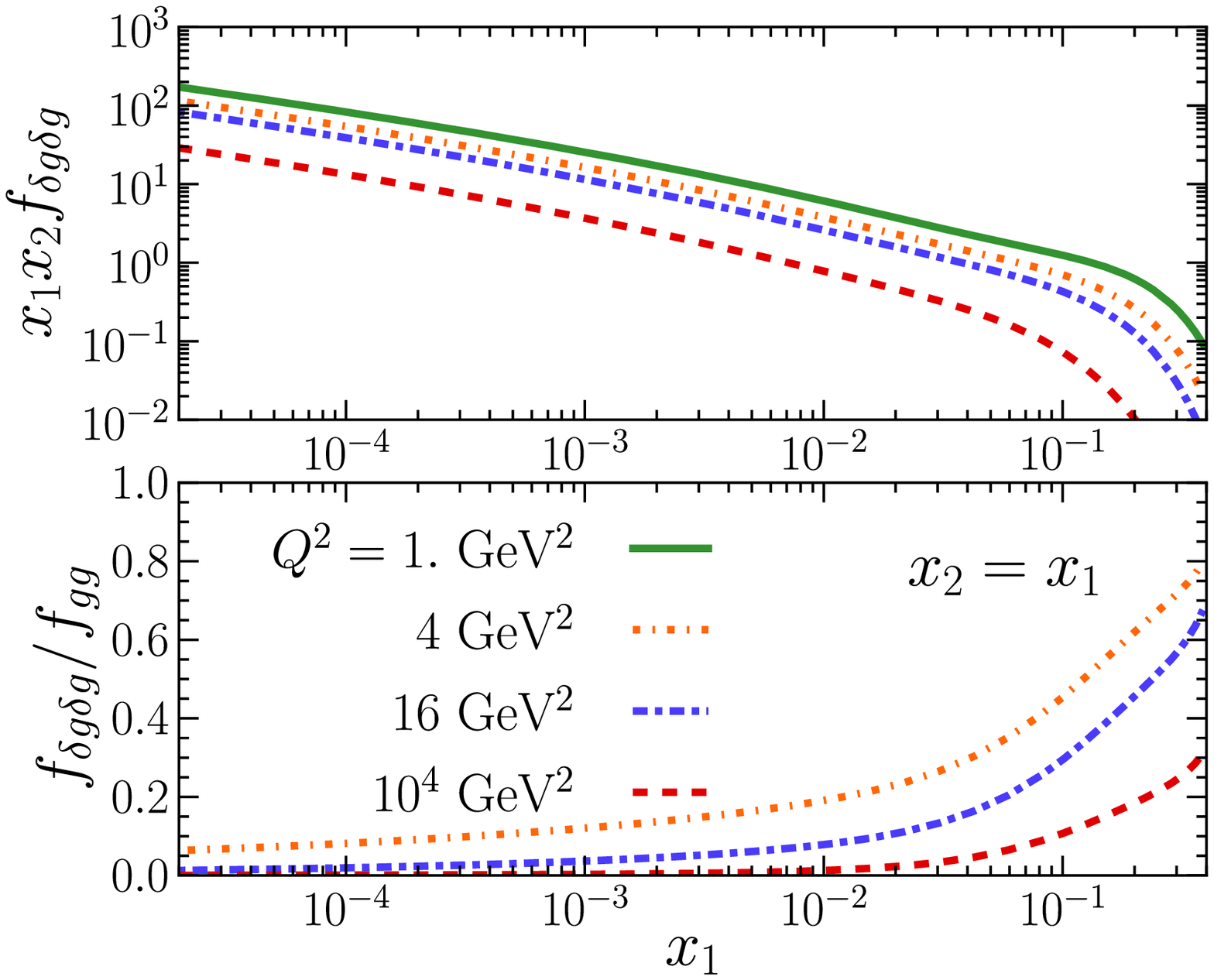} \;
\includegraphics[width=0.48\textwidth]{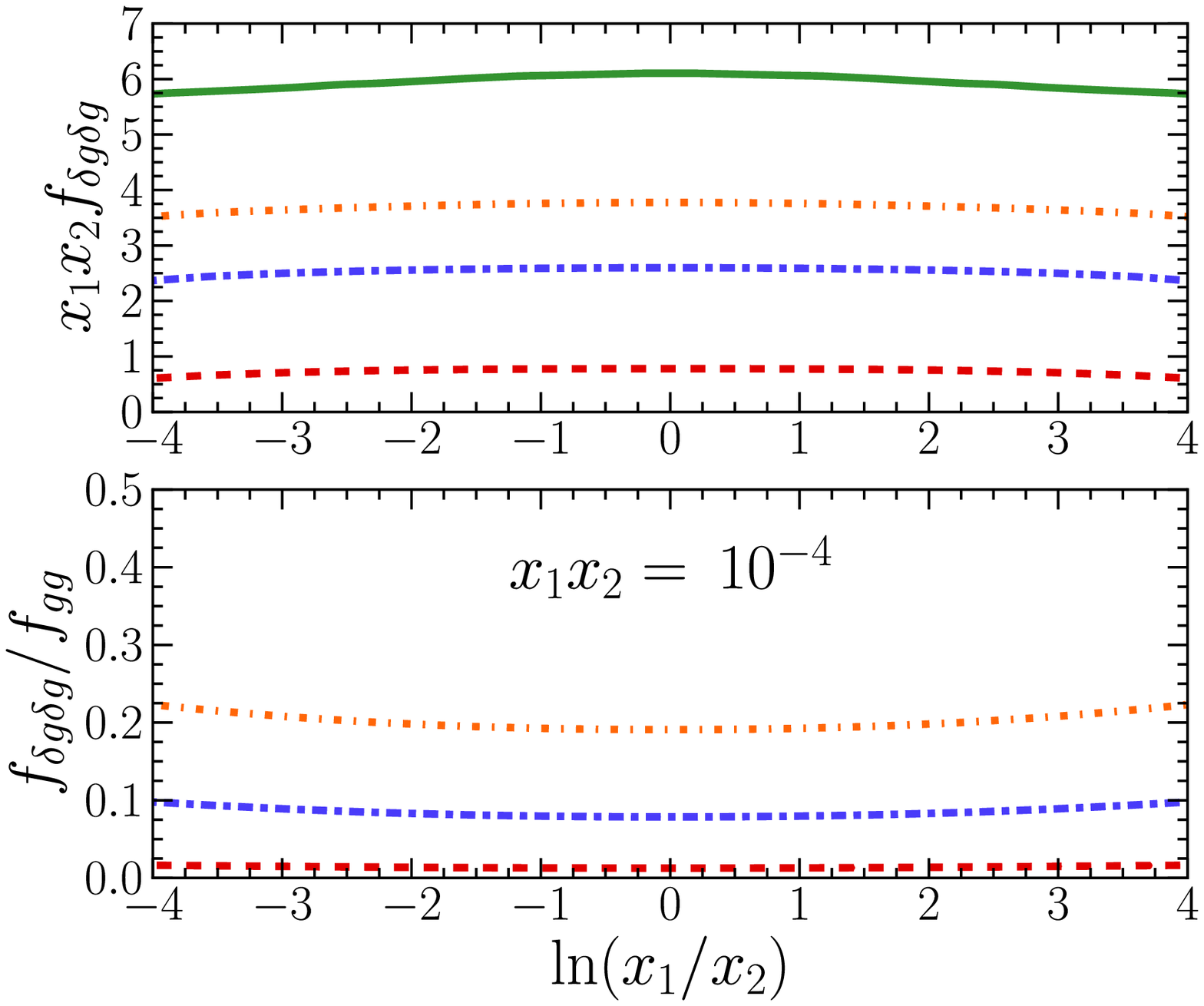}
}
\caption{Evolution of longitudinally polarized up-quarks (top) and linearly polarized gluons (bottom). Either as a function of $x_1=x_2$ (left) or as a function of $\log x_1/x_2$ (right). At the initial scale of 1~GeV, the polarization is maximized (equal to the unpolarized distribution). Lower panels show ratio of polarized over unpolarized distributions. Figure from Ref. \refcite{Diehl:2014vaa}.}
\label{fig:pol-evo}
\end{figure}

\section{Prospects}

We have seen that two-parton correlations are very interesting
properties of the non-perturbative proton structure, and they can
be relevant in {specific} DPS channels.
So far, it has been challenging to 
observe them at the LHC and 
extract dPDFs from data.
While waiting for precise data expected from LHC 
at high luminosity in the near future, 
one could look for signatures of the presence of correlations 
in an extracted quantity, 
the so-called effective 
cross-section, $\sigma_{eff}$.
Let us introduce now this quantity. 
Since dPDFs are largely unknown, 
it has been useful
to describe DPS cross section{s} independently of dPDFs, through the
approximation
\begin{eqnarray}
d\sigma^{AB}_{DPS}  \simeq \dfrac{m}{2} d\sigma_{SPS}^A \dfrac{ 
d\sigma_{SPS}^B}{\sigma_{eff}}\,,
\label{sigma_eff_exp}
\end{eqnarray}
where $d\sigma^{A(B)}_{SPS}$ is the SPS 
cross section with final 
state $A(B)$:
\begin{eqnarray}
\label{s_single}
d\sigma_{SPS}^{A(B)}= 
\sum_{i,k}  f_{i} (x_{1})
f_{k}(x_3)\,
d\hat \sigma_{ik}^{A(B)}(x_1,x_3)\,.
\end{eqnarray}
The physical meaning of Eq.~(\ref{sigma_eff_exp}) is that, 
once the process $A$ has occurred with cross section ${d}\sigma_{SPS}^A$, 
the ratio ${d}\sigma_{SPS}^B / \sigma_{eff}$ represents the probability of process $B$ 
to occur.
So far, a constant value of $\sigma_{eff}$
has been assumed in 
 the experimental analyses performed.
%,
In this way, different collaborations have extracted values of 
$\sigma_{eff}$, analyzing events with 
different final states and with different
center-of-mass energies of the hadronic 
collisions. The results have large error bars
and their central values vary 
in the range $2-20$~mb (see, for example,
Figures~8 and 9 in \cite{Aaboud:2016dea}). However, 
these numbers are to be taken with  
caution as the different extractions rely on different assumptions, 
for example, 
{with regards to the SPS} cross sections.
It is interesting
to realize that
the approximations leading to Eq. (\ref{sigma_eff_exp}),
with a constant $\sigma_{eff}$,
from Eq. (\ref{uno}), are the same leading the dPDF to
its full factorized form.
As a matter of fact,
by inserting Eqs. (\ref{fact}) and (\ref{factb})
into Eq. (\ref{uno}), 
one obtains  
$\sigma_{eff}$ from Eq. 
(\ref{sigma_eff_exp}) and (\ref{s_single}) as follows{:}
\begin{equation}
\sigma_{eff}^{-1} =  \int d{^2} \vec{y} \, 
[T(\vec{y})]^2~,
\label{bprofile}
\end{equation}
with the quantity 
\begin{equation}
T( \vec{y} )
= \int d{^2} \vec{b} \, G( \vec{b} + \vec{y} ) \, G( \vec{b} ) {\,,}
\end{equation}
controlling the
double parton interaction rate.
The fact that $\sigma_{eff}$ does not show any dependence on 
parton fractional momenta, hard scales or parton species,
is clearly a consequence of the assumptions {in Eqs.} (\ref{fact})
and (\ref{factb}).
If {those assumptions were relaxed} 
$\sigma_{eff}$ would explicitly depend on
scales and flavors, and on all momentum fractions, 
and would be a complicated average  
(with $x_i$ dependent weights) of all the correlations 
described by the double parton distributions.
One could therefore 
analyze data looking for such a dependence. 
Besides, model calculations show that  
correlations in momentum fractions
cannot be treated separately from 
those involving also {$\vec{y}$}:
the way the dPDF differs from the product of
single parton densities changes with {$\vec{y}$}
\cite{
Rinaldi:2014ddl,
Rinaldi:2016mlk}.
Using model calculations
{without the assumptions leading to Eqs. }(\ref{fact})
and Eq. (\ref{factb}),
$\sigma_{eff}$ {was} found to depend
non-trivially on longitudinal momenta.
In particular, this was obtained in the LF
constituent quark model
\cite{Rinaldi:2015cya},
as well as
in a holographic approach~\cite{Traini:2016jru}.
Very recently, the LF model calculation of dPDFs
has been used to evaluate the cross section for
same-sign $W$ boson pair production, a promising channel to look for
signatures of double parton interactions at the LHC.
In this way, the average value of the DPS cross section 
was found to be
in line with previous estimates
which make use of a constant $\sigma_{eff}$ as an external parameter, 
not necessary in this approach. The novel obtained dependence
on longitudinal momenta 
addresses the possibility to observe 
two-parton correlations, in this channel, 
in the next LHC runs \cite{Ceccopieri:2017oqe}.
An example of these results is shown in Fig.
\ref{fig:seff.eps}.

\begin{figure}
\includegraphics[width=0.75\textwidth]{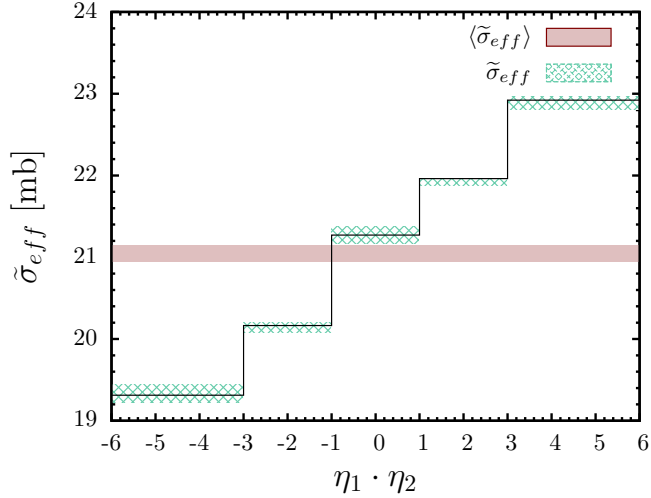}
\caption{
The quantity $\tilde \sigma_{eff} =
\dfrac{m}{2} d\sigma_{SPS}^A  
d\sigma_{SPS}^B 
/
d\sigma^{AB}_{DPS}  
$, for the production
of two $W$ bosons with the same sign,
in the kinematics of the CMS measurements
of Ref. \refcite{CMS:2017jwx}.
SPS and DPS cross sections are calculated using PDFs and dPDFs 
obtained in the LF model \cite{Rinaldi:2015cya}. No factorized structure
has been assumed for dPDFs.
In this way, a dependence on the longitudinal variable
$\eta_1 \cdot \eta_2$
is clearly predicted and could be tested
in future analyses.
$\eta_{1,2}$ are the pseudorapidities
of the detected muons in the final state, naturally related to the 
longitudinal parton momenta.
Figure from Ref. \refcite{Ceccopieri:2017oqe}, where further
details can be found.
}
\label{fig:seff.eps}
\end{figure}

Since in the DPS cross section the dependence upon
$\vec{y}$ is integrated {over},
a direct test of the breaking of Eq. (\ref{factb}) in DPS,
addressing correlations between $\vec{y}$ 
and $x_1, x_2$ in dPDFs, appears difficult at the moment.
An indirect test of these correlations is expected
from future measurements at Jefferson Lab, COMPASS and at a possible
future electron-ion collider
\cite{Accardi:2012qut}, where at least
a detailed picture of the one-body distribution
${F}_a(x,\vec{b})$,  should be at hand.

As for the correlations {between} quantum numbers described
in the previous section,
their impact {on} cross sections has been studied only in 
a limited number of cases. For the production of two $D{^0}$ {mesons}, 
{as} measured by LHCb \cite{Aaij:2012dz}, 
{the low masses of the final states allows for a large impact} 
on the size of the cross section 
from longitudinally polarized gluons, 
reaching a contribution of up to 
50\% of the unpolarized \cite{Echevarria:2015ufa}.
This is an example of the importance of further
exploratory studies of
DPS 
to find channels and phase space regions
in which two-parton correlations are more pronounced
and easily measured.
In this sense, {input} is expected also from proton-nucleus
scattering, where the DPS contribution is known to be enhanced
\cite{Strikman:2001gz}.

There are several
elements working together to provide a promising 
near future for DPS in general, and measurement of correlations in 
particular. 1) The 
continuous refinements of the DPS theory, including for example a 
scheme to combine, without double-counting, the SPS and DPS cross sections 
described in Ref. \refcite{Diehl:2017wew}, 
2) The increasing integrated luminosity collected by the experiments 
at the LHC and 3) The improved precision to which the SPS cross sections 
are know{n}.  Combined, this provides good reasons to further develop 
the theory for DPS, motivation for phenomenological studies 
of the effects correlations have on actual observables, and good 
prospects for interesting experimental results to confront the theory 
with in the upcoming years.

Double TMDs enter cross sections when the transverse momenta of
for example two vector bosons are measured and small. 
In this region, {there is} no factorization theorem
without considering both single and double parton scattering. 
The formalism to treat this region in both single and double parton 
scattering \cite{Buffing:2017mqm}, will allow for interesting prospects 
to investigate the correlations in DPS, including those between
the transverse momenta of the two partons.
{The experimental searches have now measured same-sign double-$W$ production \cite{CMS:2017jwx},}
often put forward as the 
cleanest signal for DPS.
{Interesting} results are expected 
also in channels 
where the separation between single and double parton scattering is 
less straightforward. An increased precision on both DPS 
and SPS sides
will lead to a situation where the double parton distributions are 
the {main} unknown. Using differential calculations 
and resummation at high logarithmic accuracy, for example in double 
boson production, the combination of DPS and SPS will be important 
and comparisons to data will enable 
extractions of dPDFs and interparton correlations,
or experimentally constrain them.

In summary, the increased luminosity will allow for more differential measurements. Moving towards a theory that allow for more complete phenomenological explorations, simultaneously treating both SPS and DPS, provides the basis for our belief that inter-parton correlations might soon be an experimentally established fact, or a heavily constrained hypothesis. 

\section*{Acknowledgements}
Many discussions with 
F.A. Ceccopieri, L. Fan\`o, J. Gaunt, M. Rinaldi, M. Traini, D. Treleani, V. Vento
are gratefully acknowledged. We thank S. Cotogno and T. van Daal for useful comments on the manuscript. Special thanks go to J. Gaunt and P. Bartalini for all their efforts as editors of this book. TK acknowledges support from the Alexander von Humboldt Foundation and the European Community under the "Ideas" program QWORK (contract 320389).}

\bibliographystyle{JHEP}
\bibliography{dps_corr_2}

\printindex
\end{document}